\documentclass[%
 reprint,
 amsmath,amssymb,
 aps,
prb,
]{revtex4-2}

\usepackage{graphicx}
\usepackage{dcolumn}
\usepackage{bm}

\begin{document}

\preprint{APS/123-QED}

\title{Metastable states of 2D-material-on-metal-islands structures \\ revealed by thermal cycling}

\author{V.A. Ievleva}
\author{V.A. Prudkoglyad}
\author{L.A. Morgun}
\author{A.Yu. Kuntsevich}

\email{alexkun@lebedev.ru}

\affiliation{P.N. Lebedev Physical Institute of the Russian Academy of Sciences, 119991 Moscow, Russia
}
\affiliation{HSE University, Moscow 101000, Russia}

\date{\today}

\begin{abstract}
The integration of 2D materials with artificially textured substrates offers exceptional opportunities for engineering novel functional devices.
A straightforward technological route towards such devices is a mechanical dry or wet transfer of 2D layer or heterostructure onto prepared patterned elements with subsequent van der Waals bonding. An issue of van der Waals bond stability is crucial for device operation but is almost unexplored. In our research we address it by studying transport properties of hBN/graphene heterostructures transferred onto metallic island arrays and subjected to thermal cycling. We reveal that heating from cryogenic to room temperature and cooling back leads to irreversible changes in electronic transport properties: the contact between metal and graphene degrades, and signatures of suspended
graphene regions transport disappear. These changes are accompanied by slight movement of the flakes and atomic-force-microscope-detected breakdown of van der Waals bonds between the flake and substrate near the metal electrodes.
Interestingly, a hot pressing allows to restore the metal-to-graphene contact. We relate the observed metastability to the thermal-expansion-driven flake delamination and argue that it is accompanied by redistribution of the interfacial water or organic residues. Our findings provide useful insights into the topic of interfacial stability in van der Waals heterostructures and establish constraints for low-temperature applications of transferred 2D devices. We also add up an additional control parameter for the experimentalists in the field of 2D materials - degree of quenched disorder.
\end{abstract}

\maketitle


\section{\label{sec:intro} Introduction}

Two-dimensional (2D) materials and van der Waals heterostructures (VdWH) made  from them have extraordinary and highly customizable properties. For instance, transition metal dichalcogenides possess remarkable optical characteristics\cite{mak2016photonics} and are considered as materials for transistor\cite{liu2021promises}; graphene serves as an infrared, plasmonic material and platform for novel low-energy physics\cite{yankowitz2019van}; some other 2D materials demonstrate topological\cite{zhang2022two}, memristive\cite{zhao2020current}, superconducting\cite{qiu2021recent}, magnetic behavior\cite{gibertini2019magnetic}. 

Mechanical transfer of individual layers enables the integration of vastly dissimilar materials, in contrast to epitaxial growth, which requires lattice-matched substrates and chemically compatible compounds. A particularly promising approach involves placing 2D materials or heterostructures on textured substrates or patterned electrodes, which leads to novel mechanical\cite{li2017role,shi2016tribological,yilbas2017effect}, electrical\cite{calleja2015spatial}, magnetic\cite{gao2021graphene}, optical\cite{ho2024finite,zhang2017graphene},  thermosensitive\cite{pawlak2016fully}, chemosensitive\cite{zhu2017room,holicky2024fabrication}, photodetective\cite{wang2017graphene}, and energy-harvesting\cite{khan2022cvd} properties. 
While many 2D materials degrade rapidly under ambient conditions, certain compounds such as transition metal disulfides, graphene, and hexagonal boron nitride (hBN) exhibit notable stability. 
However, it is not sufficient to  use stable materials in order to get a reliably working 2D electronic device. Durability under variable environments is another critical factor for practical applications, yet  it remains not completely understood for 2D materials and heterostructures. 

 The stability issue has been deliberately investigated for twisted 2D material-based structures\cite{silva2020exploring, liu2022twisted}. Indeed, the ultra-low friction coefficient makes sliding of one layer with respect to another possible. This phenomenon is called superlubricity\cite{dienwiebel2004superlubricity, zheng2008self, li2017superlubricity}. This low friction is typical for perfectly flat van der Waals material surfaces and is affected by the commensurability of the layers\cite{dong2019friction, liao2022uitra}. 

 The edges and local defects increase friction\cite{berman2018approaches}. Moreover, the static friction between VdW and non-VdW materials is generally higher\cite{paolicelli2015nanoscale, wang2023effects, kamenskaya2024strain, yang2025cooling}. The friction coefficient between a 2D material and the substrate is affected by the gas environment\cite{zaidi2025frictional}, moisture\cite{lee2017enhancement, yang2018roles} and organic contaminants\cite{chen2020origin}. Friction strongly depends on the number of layers and has a tendency to vary with time under load\cite{wang2017degradation, curry2021structurally}.  A textured substrate generally increases the friction\cite{li2010substrate}. However, a high enough static friction at room temperature is not sufficient to provide stability of the structure under variable temperature. 
There are limited investigations of thermal cycling in composite graphite-based materials\cite{qu2022effects,andrew2025effect}. Concerning 2D materials on substrates, a trivial yet important observation of local strain induced by the difference in thermal expansion coefficients was uncovered only recently\cite{yang2025cooling}. Such factors as local strain and thermal expansion during the thermal cycling remain unexplored for the flakes placed on textured surfaces.

In this work, we investigate the thermal cycling sustainability of graphene/hBN field effect structures transferred onto metallic electrodes. Through repeated cryogenic experiments, we demonstrate that metastability is a generic property of such systems. After the thermal cycling the low temperature conductive properties change in a definite way indicating the contamination of the graphene bottom interface and loss of the direct van der Waals contact between the graphene and the substrate. We show that this contact might be restored by a certain thermal treatment. Our findings not only impose constraints on potential device applications, but also suggest that thermally induced phase transitions in 2D-on-textured-substrate systems could lead to novel physical states.

\section{Materials and Methods}

\begin{figure*}
    \centering
    \includegraphics[width=\linewidth]{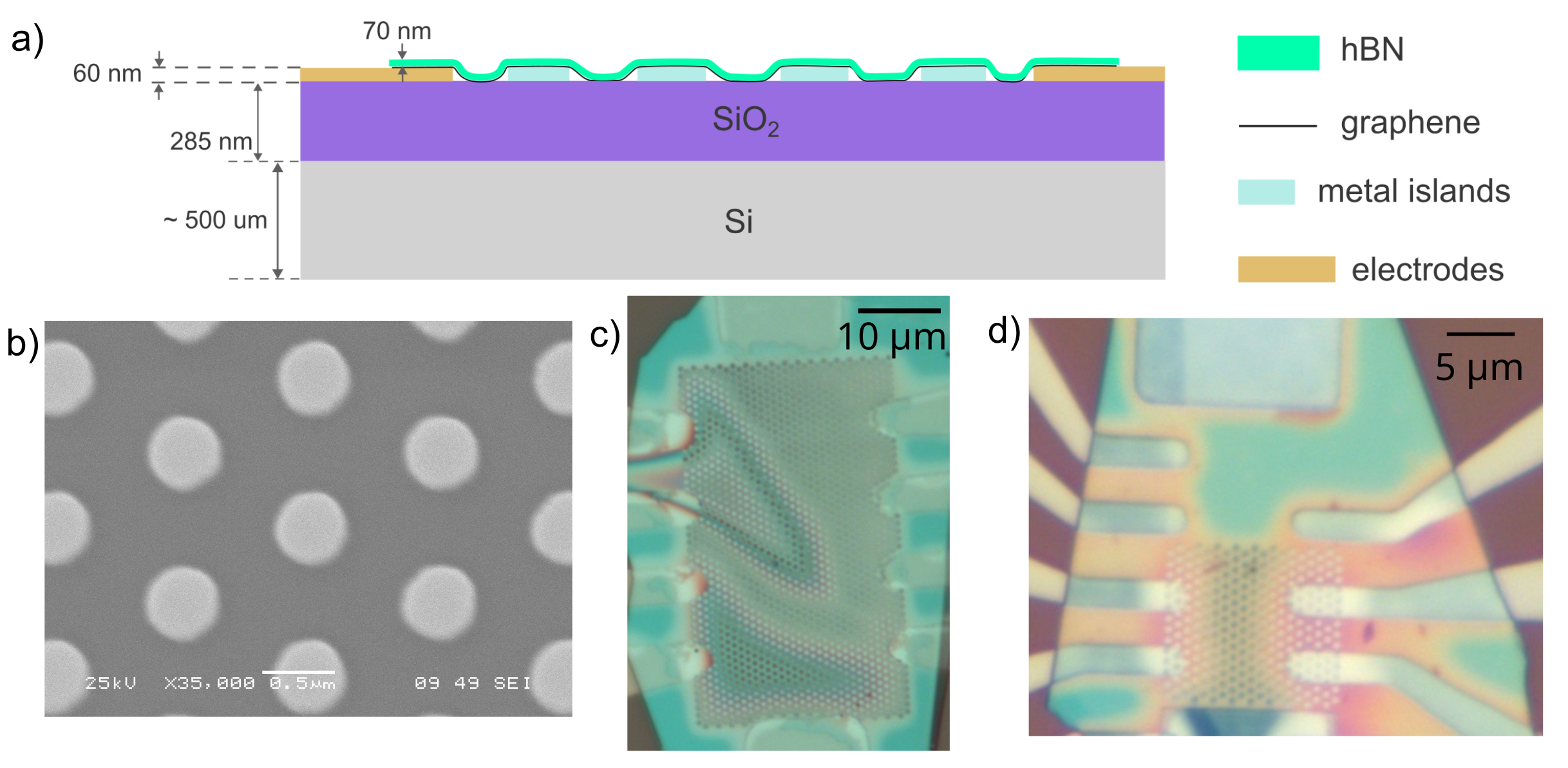}
    \caption{The samples. (\textbf{a}) Schematics of the vertical cross-section of the samples under study. (\textbf{b}) SEM image of a triangular lattice of Re islands. The same geometry was used for Nb/Pt islands. (\textbf{c}) First (Nb/Pt) sample.  (\textbf{d}) Second (Re) sample before top-gate electrode patterning. Two areas are seen: bare graphene (top) and graphene on islands (bottom).}
    \label{fig:both_samples}
\end{figure*}

\subsection{Preamble. Choice of the structures design.}
We observed the metastability on multiple samples with graphene/hBN heterostructure placed above the contacts ("bottom contact" geometry) and never observed it in the 
structures, where metal contacts were evaporated atop graphene ("top contact" geometry). Top contact geometry is frequently used; there are hundreds of publications where samples of this geometry survived several cooldowns and none of them reports metastability. Nevertheless,  the bottom contact design is considered a promising approach for graphene electronics\cite{bottomelectrode}, and we are not aware of multiple cooldown studies for the bottom geometry. 
We placed graphene/hBN heterostructure onto sub-micrometer metallic islands and contact electrodes  (see Fig.~\ref{fig:both_samples}a), to enable 4-terminal measurements.
In this paper we present the data for two representative samples studied in most detail.
A similar geometry was  used in  Refs.\cite{Bouchiat_2009,Allain2012-sx,Han2014,Ti/Al_on_G,Sun2018,Gupta_2024} where Josephson effect was studied, though the metal was  deposited on top. In our samples we do not observe the Josephson effect (in the first sample due to oxidation of the superconductor; and in the second due to insufficiently low temperatures). However, we do observe metastability!

\subsection{Sample fabrication}
Triangular lattices with a 1$\mu$m period of the 500 nm diameter discs were patterned using a double-layer (PMMA/MMA) electron beam lithography on p-doped 285 nm Si/SiO$_2$ substrates,  which were preliminary cleaned with acetone/isopropanol/deionized water and allow bottom gating. We then e-beam evaporated 50 nm thick metallic layers and performed a lift-off.  For evaporation we used Plassys MEB 550 and 400; residual gas pressure before the evaporation was at least 5*10$^{-8}$ mbar low, no specific substrate heating or ion beam cleaning was used.  In evaporation we used two different metals: a Nb with a 4 nm thick Pt top layer (first samle) and amorphous rhenium\cite{tarkaeva2025high} (second sample). An image of a metal disc array is shown in Fig. \ref{fig:both_samples}b. Using photolithography we formed $\sim$70 nm thick contact electrodes of Cr/Au for the Nb/Pt sample and Ti/Pt for the Re sample. A heterostructure  consisting of a mechanically exfoliated monolayer graphene and a 70 nm thick hBN flake  was assembled and transferred using a hot dry method with a PPC-covered PDMS drop \cite{martanov2020} atop of the array of islands and contacts. Sample images are shown in Figs. \ref{fig:both_samples}c,d.  Additionally, we fabricated a top gate ($\sim$ 200 nm thick, e-beam evaporated aluminum) in the Re sample. In Fig. \ref{fig:both_samples}, d we present its image without top gate for clarity, further in Fig. \ref{AFMscans}-\ref{fig:raman} the top gate is represented. 

\subsection{Characterization}
Several  techniques were used to characterize the samples and the 2D materials during the fabrication. Atomic force microscopy in tapping mode (NT-MDT Solver) was used to determine hBN thikcness and morphology of the structure. Scanning electron microscopes (Helios Nanolab 660 FEI and JEOL JSM-6460) were used to take images of the structures and produce electron-beam lithography. Monolayer graphene was selected using Raman spectroscopy.  To measure Raman spectra we used an express-analyzer operating at wavelength of 532 nm (EnSpectr R532) combined with Olympus BX51 metallographic microscope. We used 50x objective witn NA=0.8

\subsection{Low temperature measurements}
Transport properties of the samples were measured using 4-terminal scheme at a transport current low enough to avoid overheating (below 100 nA).  Lock-in detection at frequencies between 10 and 313 Hz was employed in various cryo-magnetic systems including a BlueFors dilution refrigerator, a Cryogenic 0.3K liquid He3 cryostat, a Cryogenic CFMS  He4 cryostat.  In the dilution refrigerator the sample was held in vacuum, with the thermal contact to the dilution chamber provided through the Cu plate, while in the He4 or He3 cryostats the sample was placed in vapor. The cooling rates from room temperature to 4K were 0.15 K/min, 1 K/min and 2.5 K/min in the dilution, He4 and He3 refrigerators, respectively. We provided one cryogenic cycle from room temperature to cryogenic temperatures in each experiment described below. For magnetic field dependencies of the resistivity/Hall resistivity the field was swept from positive to negative values and the data were symmetrized/antisymmetrized, respectively.

\section{Results}

\subsection{First sample}
\begin{figure*}
    \centering
    \includegraphics[width=\linewidth]{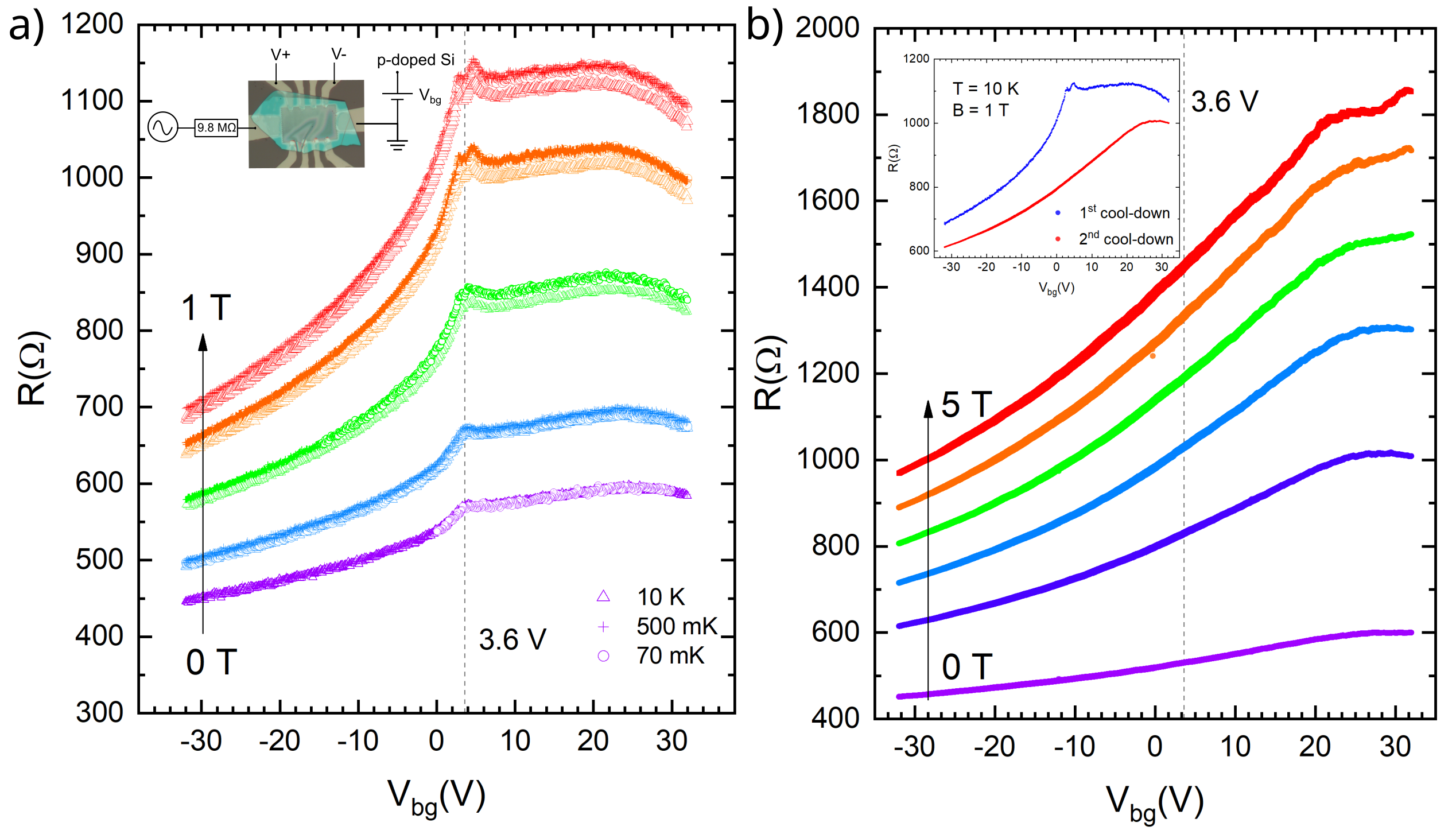}
    \caption{First (Nb/Pt) sample. (\textbf{a}) Gate voltage dependencies of the resistance in perpendicular magnetic field obtained during the first cooldown. Magnetic field amplitudes consequently from violet to red are 0, 0.2, 0.5, 0.8, 1 T. For each color there are three almost coincident data sets corresponding to three temperatures: 70 mK (circles), 500 mK (crosses), 10 K (triangles). Inset: four-point measurement scheme. (\textbf{b}) Results of the same measurements after the second cooldown. Magnetic field amplitudes from violet to red are 0, 1, 2, 3, 4, 5 T. The measurements are taken at 10 K. Inset: a direct comparison of the dependencies obtained in a first and in a second cooldown at B = 1 T, T = 10 K.}
    \label{fig:firstsample}
\end{figure*}

Fig.~\ref{fig:firstsample}a shows the gate voltage dependence of the Nb/Pt sample resistance  during the first cooldown. At zero magnetic field, independently of temperature two resistance peaks  are observed: a shallow maximum for $V_{g2}\approx 20$ V and a feature at $V_{g1}\approx 3.6$V. For non-zero perpendicular magnetic field $B$ the feature evolves into a small sharp peak, which splits into two in $B>0.5$~T.  

In order to interpret the data, one should consider graphene on the islands as an effective medium consisting of three components: graphene on metal, graphene on SiO$_2$ substrate, and suspended graphene. Suspended regions arise because a graphene/hBN heterostructure due to its finite rigidity cannot cover tightly all the metallic structures on a substrate (as shown in Fig. \ref{fig:both_samples}a).

The conductivity of such a medium should be a combination of the conductivities of its components. We may use the following qualitative considerations. Graphene on the islands is supposed to have no gate voltage dependence of the resistance due to electric field screening by the metal. It could be either short-circuited by the metal or not, depending on the contact resistance. In this particular sample we believe the contact resistance is rather high (see comparison with the second sample below). Graphene parts suspended near the island and contact edges are believed to be the cleanest ones \cite{bolotin2008ultrahigh,dai2023suspended} and have a charge neutrality point (CNP) at low-$Vg$, seen as the small peak. Graphene on a SiO$_2$ substrate gives a shallow peak far from zero gate voltage, typical for hole-doped graphene. Magnetic field usually induces positive magnetoresistance in graphene independently of gate voltage\cite{friedman2010, gopinadhan2015}. More interestingly, low-$V_g$ resistivity peak splits as  the magnetic field increases. This splitting does not depend on temperature at low $T$ and hence could be explained in the following classical way.

Magnetoresistance in graphene is the highest at the CNP\cite{cho2008charge}; therefore, the resistance  peak at the CNP further sharpens in a perpendicular magnetic field. Apparently,  the low-V$_g$ feature seen at zero magnetic field, was actually a sum of sloping peaks originating from parts of the sample with slightly different properties. As magnetic field increased, the two  constituent peaks became distinguishable. The relative sharpness of the split peaks, compared to the shallow maximum at high gate voltages, confirms the origin of the low-$V_g$ feature from the high-mobility suspended graphene regions.

 After heating up to room temperature, the sample was transferred to a different cryostat and cooled down again. In Fig.~\ref{fig:firstsample}b, one can see that the low-$V_g$ features disappeared (\ref{fig:firstsample}b, inset)! This observation means that the suspended regions lost their high mobility. Due to the rigidity of the graphene-hBN heterostructure it is hard to suspect that suspended graphene came into contact with the SiO$_2$ during the second cooldown. A more realistic scenario is that the suspended parts become contaminated and the mobility dropped.

\begin{figure}[h!]
    \centering
    \includegraphics[width=0.9\linewidth]{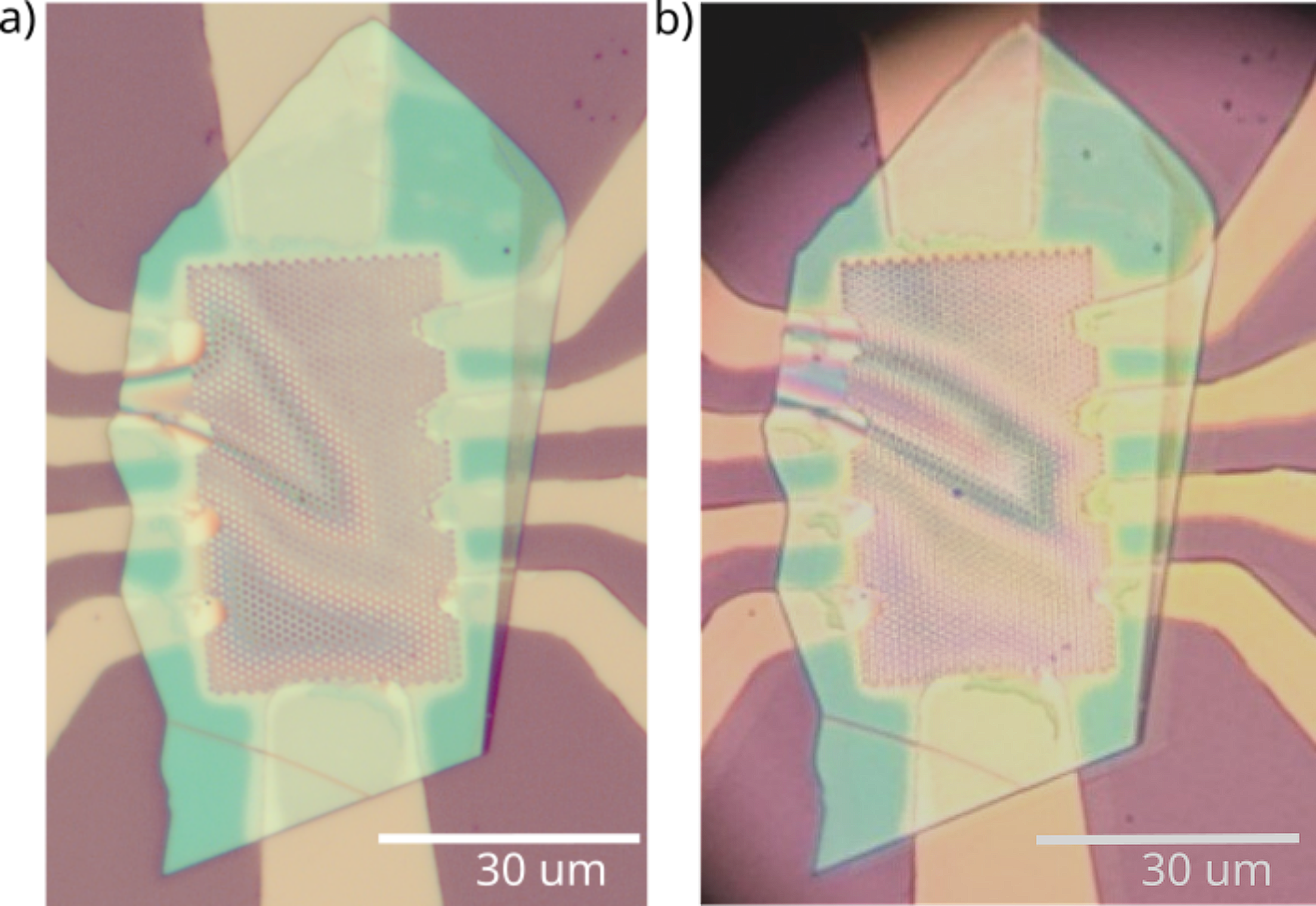}
    \caption{Visual changes in the Nb/Pt sample. A microphotograph of the sample a) before and b) after the first cryogenic cycle of measurements.}
    \label{visual_change}
\end{figure}

 Optical microscope images in Fig.~\ref{visual_change} clearly  indicate the changes. The sample was not uniform from the very beginning and a bubble in the hBN was clearly seen. This bubble appeared during the annealing, indicating that the van der Waals bond between the array of metallic islands and the graphene/hBN heterostructure is rather weak. After the cooldown, the bubble changed its shape. Careful comparison of the images in Fig.~\ref{visual_change}a and Fig.~\ref{visual_change}b reveals numerous differences across the array of metal islands and near-contact regions. These changes signify that mechanical motion of the heterostructure takes place. This mechanical motion could be related to the disappearance of the low gate voltage feature.

\subsection{Second sample}
\begin{figure*}
    \centering
    \includegraphics[width=0.9\linewidth]{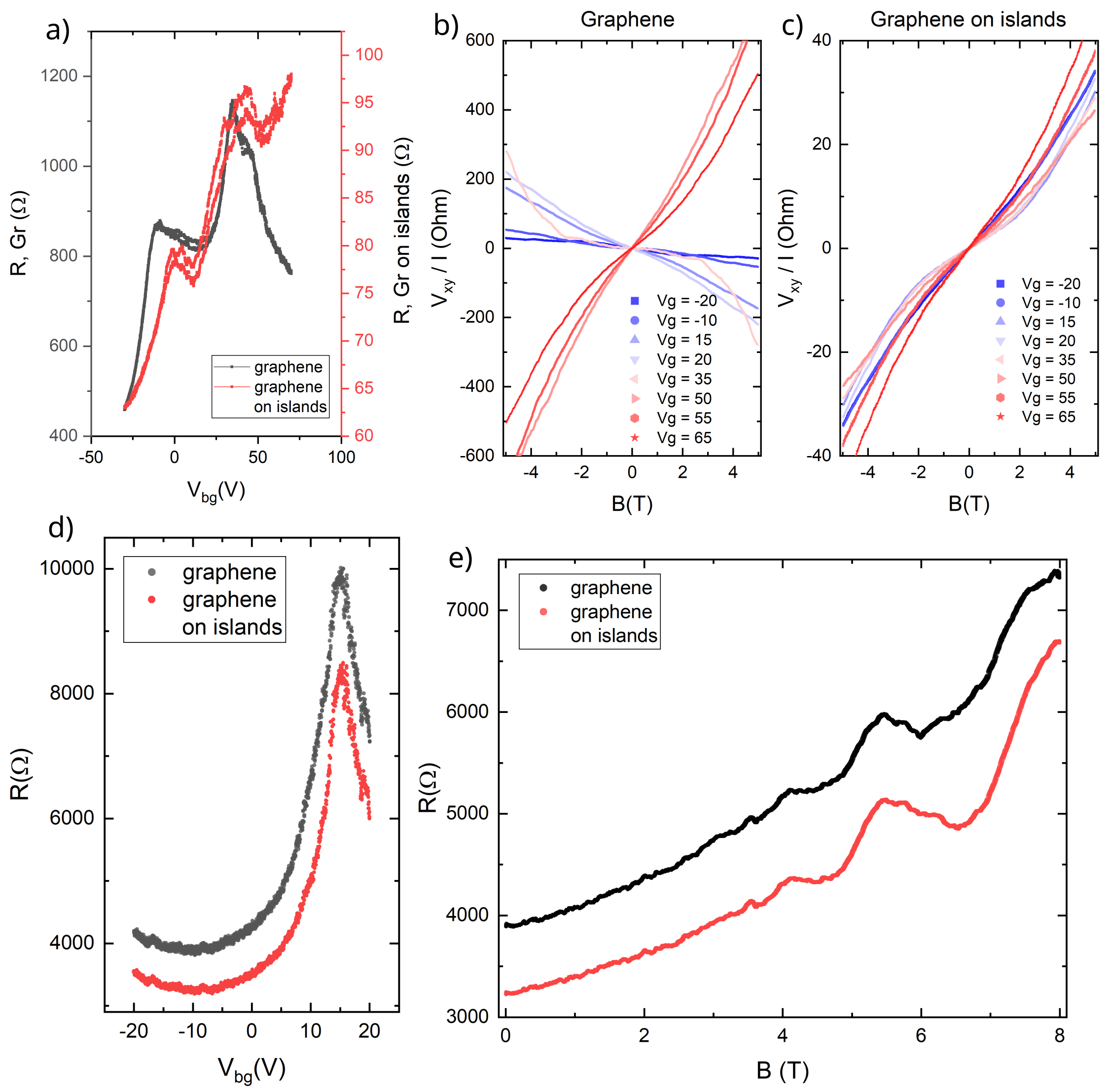}
    \caption{Low temperature properties of graphene-on-Re sample. (\textbf{a-c})- first cooldown, (\textbf{d-e}) - second cooldown: (\textbf{a})  $R(V_g)$ dependencies of bare graphene (black) and graphene on islands (red) at T = 10 K; the Hall resistance dependence on magnetic fields at different gate voltages at T = 10 K of bare graphene (\textbf{b}) and graphene on islands (\textbf{c}); (\textbf{d}) $R(V_g)$ dependencies and (\textbf{e}) the magnetoresistance oscillations in perpendicular magnetic field at $V_g = -10 V$ for bare graphene (black) and graphene on islands (red) at T = 0.3 K.}
    \label{fig:secondsample_1-2_CD}
\end{figure*}

The second, rhenium-based sample during the first cooldown also demonstrated a two-peak $R(V_g)$ dependence,  which was different in details from the Nb/Pt Sample. To better understand the transport properties of the structure we intentionally fabricated two regions in series: bare graphene and graphene on metallic islands. These two areas have different transport characteristics: the resistivity of the graphene on rhenium is an order of magnitude smaller, as shown in Fig.~\ref{fig:secondsample_1-2_CD}a, because the Re islands short-circuit and partially dope the graphene. Yet, $R(V_g)$ dependence for graphene on islands   exhibits the same features as that of the bare graphene, clearly identifying two regions: suspended graphene and graphene on the substrate. Graphene without an array of islands has suspended regions located near the potential electrodes. 
As one can see in Fig.~\ref{fig:both_samples}d,  the geometry of the Re sample is such that the contacts occupy a significant area of the sample. The near-contact areas are responsible for the low-$V_g$ CNPs on the black curve in Fig.~\ref{fig:secondsample_1-2_CD}a.

   The conductance between the graphene and the islands is high, as confirmed by the Hall effect data presented in Fig.~\ref{fig:secondsample_1-2_CD}b-c. Indeed, the low-temperature Hall coefficient for the bare graphene is much larger and changes its sign with the gate voltage, while the Hall coefficient  for the graphene on islands is an order of magnitude smaller, does not change sign and does not depend strongly on the gate voltage. Qualitatively, such behavior is clear: the metal parts have low mobility and extremely high carrier density, leading to a decrease in the Hall coefficient. It is potentially possible to fit quantitatively the nonlinear Hall data using analytical theory\cite{Kuntsevich2020}; however, such fit requires too many adjustable parameters. Therefore, in this paper, we concentrate on the metastability of the properties.

Similarly to the Nb/Pt sample, the Re sample was warmed to room temperature and cooled down for the second time. The $R(V_g)$ dependencies for the graphene and graphene-on-islands  region became nearly  identical and  exhibit a single CNP at an elevated gate voltage, as shown in Fig.~\ref{fig:secondsample_1-2_CD}d. The resistance of the graphene on islands increased more than by an order of magnitude, signifying the loss of the electrical contact between the Re islands and the graphene. In a perpendicular magnetic field magnetoresistance oscillations appeared (Fig.~\ref{fig:secondsample_1-2_CD}e) with the similar parameters for both parts of the sample, showing that 
the system had become much more uniform. The preservation of almost the same resistivity values in the first (Nb/Pt) sample after the second cooldown suggests that the electrical contacts to the metallic islands there were poor from the very beginning.

 The loss of electrical contact between the graphene and the islands after cooldown and warm-up means that the van-der Waals bonds are  disrupted. In the Nb/Pt sample the clean graphene, tensed between the islands and the substrate, lost its high mobility after warming to room temperature. In the Re sample during the second cooldown the graphene lost electrical connection to the islands. These observations suggest that a very thin layer of water or organic contaminant intercalates between the graphene bottom surface and the substrate. Such a layer would inevitably be present at almost any surface  under ambient conditions.

 \subsection{AFM signatures of the metastable state}
\begin{figure}[h!]
    \centering
    \includegraphics[width=0.8\linewidth]{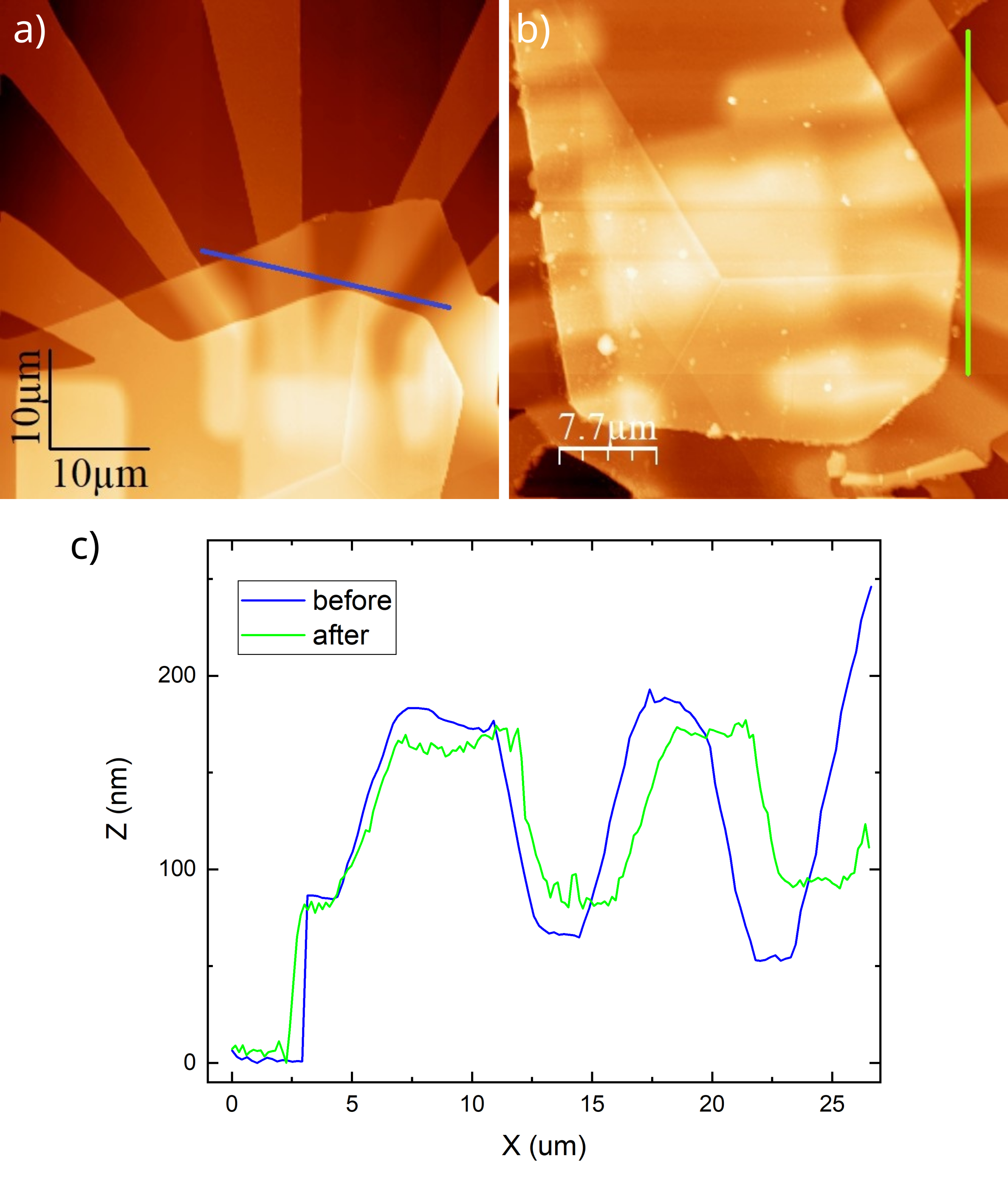}
    \caption{AFM scans of the graphene-on-Re sample before and after the first cooldown - panels (\textbf{a}) and (\textbf{b}), respectively. Panel (\textbf{c}) shows the compararison cross-section along the blue and green lines, shown in panels (\textbf{a}) and (\textbf{b}), respectively.}
    \label{AFMscans}
\end{figure}

To explicitly demonstrate the changes caused by thermal cycling, we performed atomic force microscopy scans in tapping mode on the second sample before and after the first cooldown (Fig. \ref{AFMscans}(\textbf{a,b})). 

The finite radius of the AFM tip and the presence of the top gate electrode make the array of islands region non-informative. 
We therefore compare the near-contact regions, which is also shown to be a metastable region. It can be seen from Fig. \ref{AFMscans}(\textbf{c}) that the hight of the of the stap at the contact pad edge covered with hBN decreased after the thermal cycling by several tens of nanometers. This difference could not be related to the measurement error. 
The thickness of neither hBN nor metal layer changed. One may assume that hBN layer, initially strained by the contact electrodes, lost its van der Waals bond to the substrate becoming less strained and suspended. Consequently its observed thickness decreased.

\subsection{Raman spectroscopy detection of the mechanical tensions}

\begin{figure}[h!]
    \centering
    \includegraphics[width=0.9\linewidth]{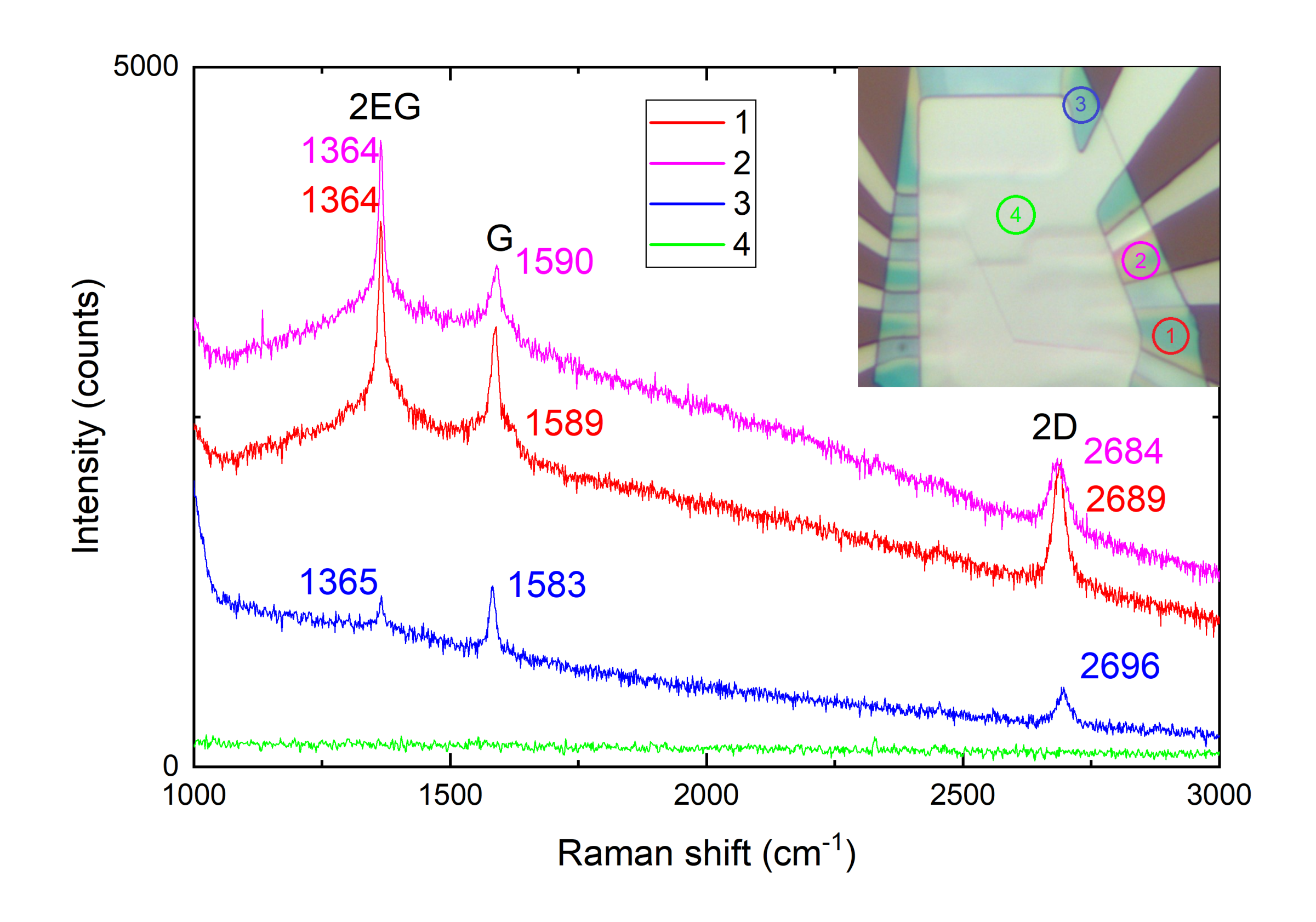}
    \caption{Raman scattering spectra of different points of the Re sample after measurements. Under top gate (green line) no signal could be collected due to the light reflection. Near the peaks their Raman shifts are written. }
    \label{fig:raman}
\end{figure}

A key element of the studied structures, that distiguish them from flat van der Waals heterostructures is the inevitable presence of mechanical strain due to bending and static friction. A direct tool to probe this strain is Raman spectroscopy. Fig.~\ref{fig:raman} shows the Raman spectra collected at different points of the Re sample after cooldown and subsequent warm up back to room temperature. Three peaks are clearly visible: E$^2_g$ peak of hBN at $\sim$1360 cm$^{-1}$, graphene G peak at $\sim$1590 cm$^{-1}$, and graphene 2D peak at $\sim$2690 cm$^{-1}$. The absence of D and D' peaks indicates high quality of the graphene. At point 3 (the blue line) we believe the influence of the metallic edge is minimal, as well as FWHM of the graphene peaks (14 cm$^{-1}$ and 25 cm$^{-1}$ for G and 2D peak, subsequently). Near dots 1 (red line) and 2 (magenta line) in the optical image more non-uniformity of hBN color can be seen, assuming greater height gradient and mechanical tensions. FWHM of the 2D peak equals to 28 cm$^{-1}$ and 39 cm$^{-1}$ for point 1 and 2.  In these areas combined effect of the tensile strain \cite{Neumann2015} (red shift and broadening of 2D peak) and doping (blue shift of G peak) is observed \cite{Bruna2014}. 

\subsection{Restoration of electrical contact by thermal cycling}
In order to affect the possibly broken van der Waals bond, we decided to 
press the second, Re-based, sample with a hot PDMS drop (150\textcelsius{} for 10 minutes {with  a force of approximately 0.5}~N) - i.e. perform the same operation as during the sample assembly. This action required detaching the sample with bonded contact wires from the measurement platform, placing it on a heating stage and returning it again with re-bonding of the contacts. The operation was risky and we are not aware of anybody having done anything similar. To avoid damaging the sample we did not clean it after the PDMS pressing. After this procedure, we cooled the Re sample down to helium temperatures and measured its transport characteristics for the third time. 

\begin{figure}[h!]
    \centering
    \includegraphics[width=0.9\linewidth]{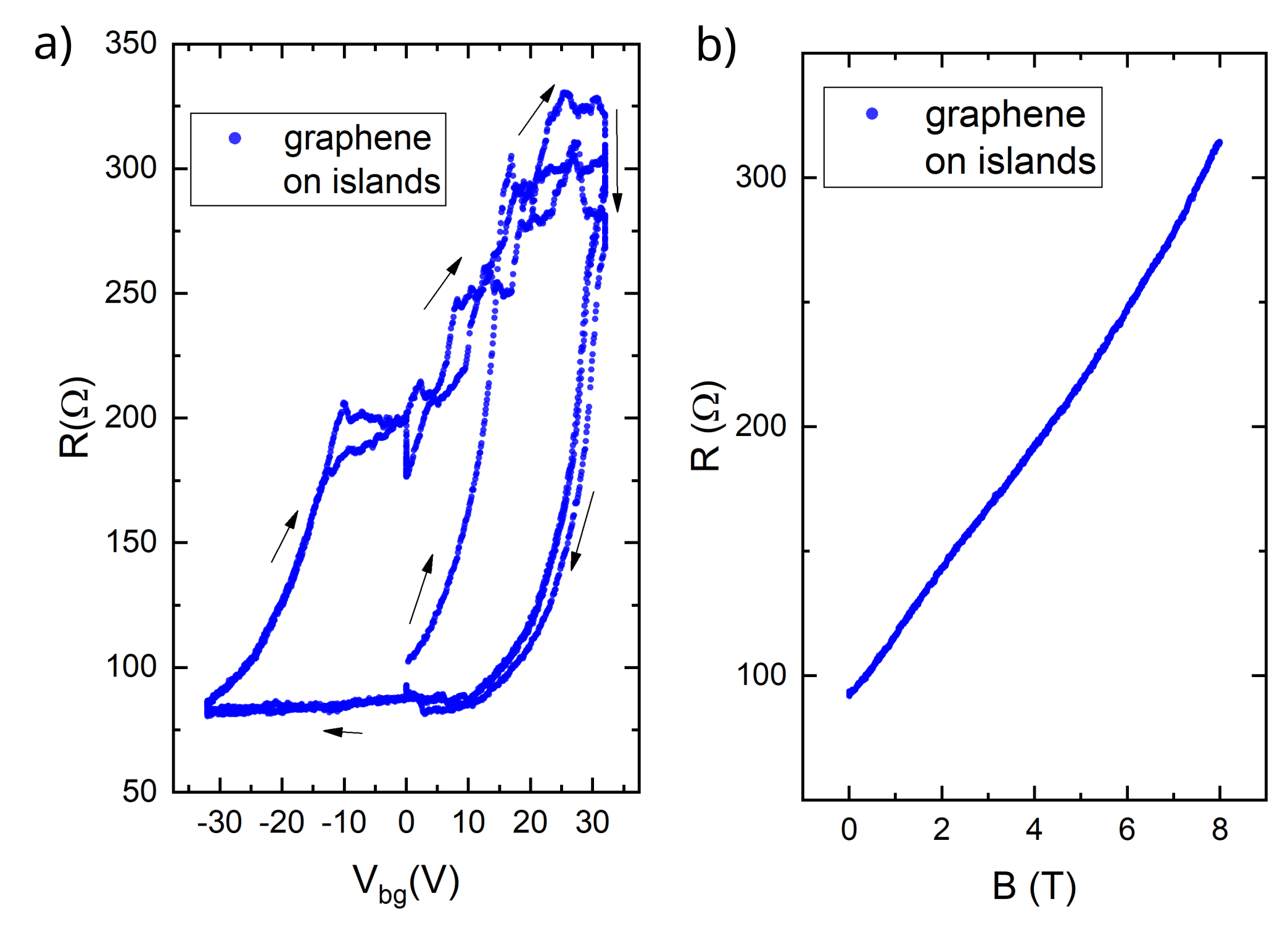}
    \caption{The third cooldown of the Re sample: (\textbf{a}) a hysteresis-like dependence of resistance on gate voltage in graphene on islands, (\textbf{b}) a dependence of resistance on perpendicular magnetic field at $V_g = -10$ V in graphene on islands. All measurements are taken at T = 0.3 K}
    \label{fig:secondsample_3_cd}
\end{figure}

Fig.~\ref{fig:secondsample_3_cd}a shows the dependence of the graphene-on-islands resistance on the gate voltage. The resistance value was restored to 100-300 $\Omega$, compared to 4000-8000 $\Omega$ in the second cooldown, which suggests the recovery of the electrical contact between the graphene and the Re islands!  
Qualitatively, the difference of the $R(V_g)$ dependence from the first cooldown (Fig.~\ref{fig:secondsample_1-2_CD}a) is not surprising for such a fragile sample. $R(V_g)$ dependence also became hysteresis-like. In Fig.~\ref{fig:secondsample_3_cd}a we plot several loops and show the direction of the gate voltage sweep. Each loop does not exactly reproduce the previous. Since the sample was not cleaned after the hot pressing we attribute this hysteresis to charge traps caused by organic contamination\cite{egginger2009current}.

In Fig.~\ref{fig:secondsample_3_cd}b, the resistance dependence on a perpendicular magnetic field at $V_g = -10$ V is shown, demonstrating the absence of the oscillations seen in Fig.~\ref{fig:secondsample_1-2_CD}e. This confirms that the islands started to be electrically connected to graphene again. Indeed, in the random potential of the islands, the graphene ceased being uniform, and was short-circuited by the islands leading to reduced magnetoresistance.

\section{Discussion}
\subsection{Wetting transition}
\begin{figure}[h!]
    \centering
    \includegraphics[width=0.9\linewidth]{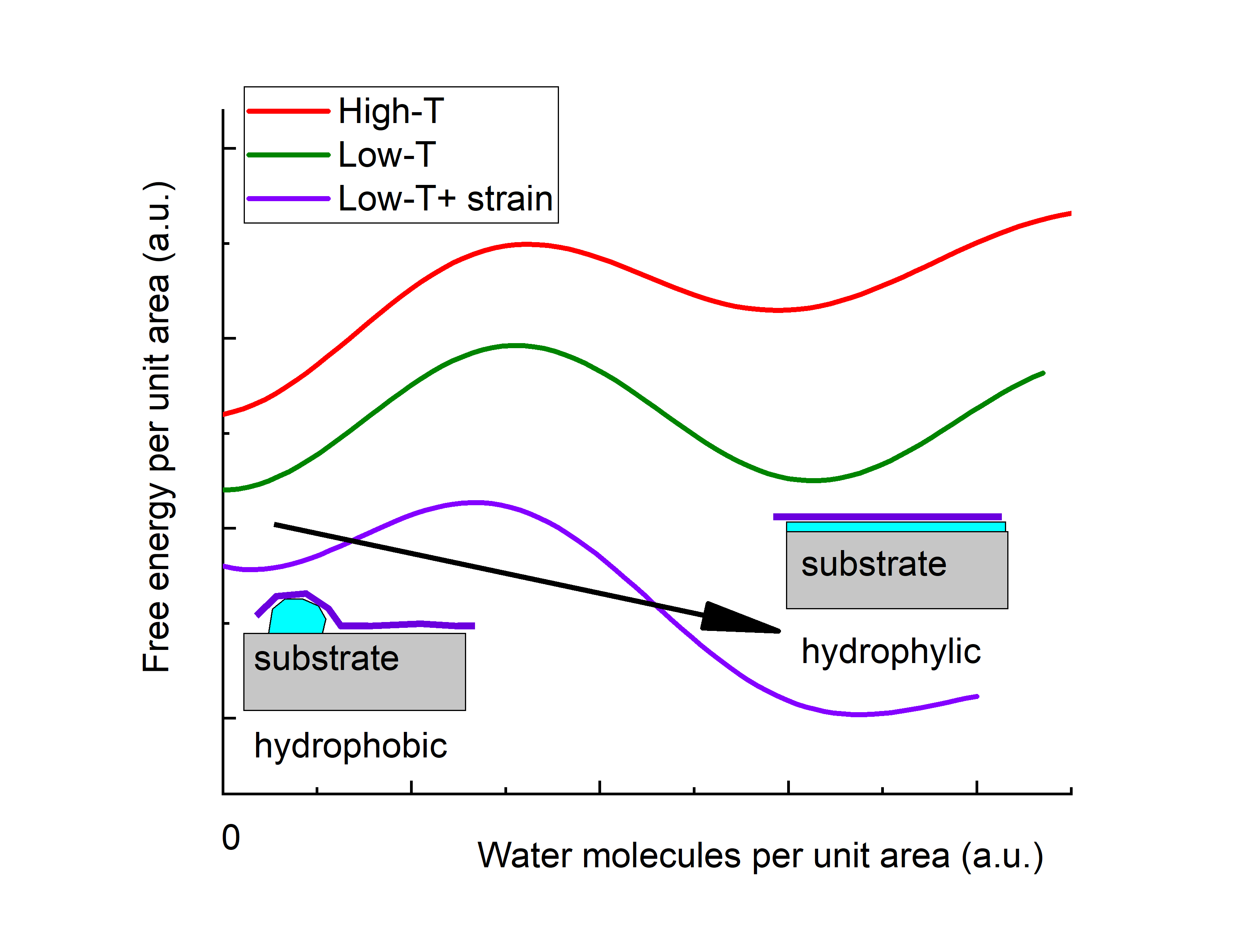}
    \caption{Free energy dependence on surface area covered with water at different temperatures.}
    \label{fig:diagram}
\end{figure}

 The redistribution of a few molecular layers of water or organic contaminants along with the first order nature of wetting-unwetting phase transition, can explain the disappearance of the charge neutrality peak at nearly zero gate voltage.
Graphene itself is hydrophilic\cite{hong2016} at room temperature unless intentionally damaged\cite{hydrophobic}. However, when pressed at room or elevated temperatures to the surface of the other van der Waals material it can undergo a hydrophobic collapse\cite{wakolbinger2020locally}, causing surface water to migrate either to the structure edges or to micro or nanometer-scale droplets. As a result, most of the interface becomes clear, including the suspended parts near the edges of the metal. This phenomenon known as self-cleansing\cite{haigh2012selfsleansing} allows for the fabrication of extremely clean VdWHs by mechanical assembly. A global minimum of the free energy in a hydrophobic state at elevated temperature is shown by the red curve in Fig.~\ref{fig:diagram}. This diagram shows the dependence of the system's free energy on the area covered with surface water. 

At lower temperatures, the interface becomes more hydrophilic owing to the decrease in the entropy factor\cite{wang2008temperature} - see the green curve in Fig.~\ref{fig:diagram}. This is a universal tendency -- all surfaces demonstrate better sorption at lower temperatures. However, when the sample is cooled down for the first time, it freezes in the hydrophobic state due to a potential barrier,  i.e., the first order nature of the wetting phase transition. We experimentally observe high mobility in the unwetted suspended regions and strong interaction between the graphene and the metallic islands.

In order to jump from one local minimum to another, van der Waals bonds must be broken and a potential barrier prevents their restoration.

Significant mechanical strain present in a heterostucture transferred onto a textured substrate (which is absent in the case of top contacts) further lowers the system
 global wetting free energy minimum(violet curve in Fig.~\ref{fig:diagram}). However, this diagram does not explain why the changes are irreversible. The system is definitely in the unwetted state at cryogenic temperatures, and becomes wetted only after heat-up. There must be a trigger for the changes. 

It can be seen from the optical microscope images in Fig.~\ref{visual_change} and confirmed by AFM measurements in Fig.~\ref{AFMscans} that the hBN heterostructure moves slightly after the thermal cycling. We believe that this slight movement dynamically triggers the wetting transition pretty much like a charged particle triggers condensation in a Wilson chamber. Once a critically high strain is formed dynamically the wetted region spreads over the entire sample.

\subsection{Possible mechanism of delamination}

\begin{figure}[h!]
    \centering
    \includegraphics[width=\linewidth]{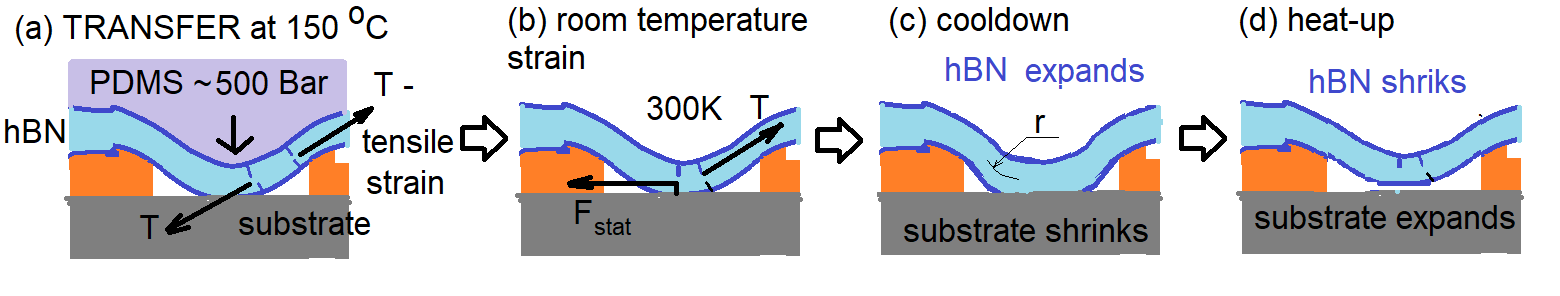}
    \caption{Schematics of van der Waals bond formation and key aspects of different stages.}
    \label{fig:delamination}
\end{figure}

To trigger wetting and move the system from one local free energy minimum to another it is crucial to break the van der Waals bonds between the substrate and the graphene/hBN heterostructure by introducing high strain. We suggest the following microscopic mechanism for this bond breaking. 
The assembly of the structure includes a hot pressing with a PDMS drop. The measured force of 0.5~N is applied to a 100$\mu$m diameter spot, so the pressure under the drop, $\Delta P$, is rather high -  about 500 Bar, see Fig.~\ref{fig:delamination}a. 

This pressure leads to an even higher Laplace strain $T$ of hBN on the order of  $\Delta P r/d$, where $r$ is the radius of curvature and $d$ is the thickness of the hBN. 
When PDMS drop is removed, the strain does not relax completely, because of the static friction force $F_{\text{stat}}$ between the flake and the substrate, see Fig. \ref{fig:delamination}b, so the structure remains strained near the points of contact with the substrate and the metal. Some other regions are free of strain. Static friction between the graphene and the metal or SiO$_2$ stabilizes the state of the structure. Then the system is then cooled down (Fig.~\ref{fig:delamination}c). 

Although the thermal expansion coefficient of the 2D heterostructures is a rather complicated matter\cite{zhang2022effect}, in our case it is dominated by the hBN layer because it is two orders of magnitude thicker than graphene. Hexagonal boron nitride is remarkable for its anisotropic thermal expansion (TE)\cite{paszkowicz2002lattice, perottoni2025thermal} with the in-plane TE coefficient $\alpha_a$ being negative and the out-of-plane coefficient $\alpha_c$ being positive. Metal, silicon and SiO$_2$ are characterized by a positive isotropic $\alpha$.
This fact is extremely important: it means that the tensile strain \textit{decreases} during the cooldown, because the metal and SiO$_2$ shrink by approximately $\sim$0.1\% while hBN expands by about 0.05\% \cite{paszkowicz2002lattice}! Taking the in-plane elastic constant of hBN, $c_{11}$, to be about 800 GPa\cite{bosak2006elasticity}, one estimates the strain reduction as relative elongation multiplied by $c_{11}$ which gives about 10~kBar. Thus, the strained part could be significantly released during the cooldown. Strain reduction leads to increase of the van der Waals contacts area, and decrease of the radius of curvature $r$ at the point of contact, as shown in Fig.~\ref{fig:delamination}c.
 
 Maximal static friction always exceeds the dynamic dry friction. When the system heats up an avalanche delamination process is initiated due to the reduced radius of curvature, and the system does not stop at the state shown in Fig.~\ref{fig:delamination}b and proceeds further to the state in Fig.~\ref{fig:delamination}d. The delamination process occurs at relatively high temperatures where the TE coefficients are large and is accompanied by wetting; i.e., system transition to the second local minimum in Fig.\ref{fig:diagram}.

The delamination of the heterostructure from the substrate is evidenced by the data in Fig.~\ref{AFMscans}c. It begins near the metal edges. The delaminated graphene bottom surface is hydrophilic\cite{hong2016} and therefore becomes covered with a water layer at relatively high temperature. Anyway the sample remains in this new minimum during the second cooldown and thereafter, as verified by multiple cooldown experiments. 
Note that the diagrams in Figs.~\ref{fig:diagram} and \ref{fig:delamination} provide only a qualitative level of physical reasoning. Quantitative calculations for such a  system are hindered by numerous unknown parameters, so we restrict ourselves to a general discourse. Importantly, our work highlights the complexity of phenomena in heterostructures based on 2D materials under thermal cycling. 

\subsection{Further development}
Our study experimentally uncovers a critical yet previously overlooked aspect of mechanically transferred 2D material structures: their metastability induced by strain, thermal expansion and likely enhanced by interfacial water layers trapped between the substrate and the 2D flakes. This effect emerges during thermal cycling.

 Notably, the effect is absent in the structures without significant strain, such as those with evaporated top contacts or 1D edge contacts. For example, in Refs.~\cite{fong2013measurement, alexander2016giant} the same sample was measured sequentially in two cryostats and the results were reproducible; in Refs. \cite{balci2012rapid, dorgan2013high} graphene with top contacts survived after powerful current annealing. Van der Waals bond with a textured surface is different: it is generally weaker and formed with strain due to static friction. Mechanical strain at the edges of metallic islands promotes delamination during the thermal cycling.

We show that despite the fabrication simplicity and promising room-temperature applications\cite{bottomelectrode}, such transferred structures require further careful optimization to achieve reproducible performance. One option is to use current annealing to stabilize the electrical response \cite{Wang2010-rm}. The  metastable state should also be responsive to laser annealing. Mechanical perturbation via atomic force microscopy at room temperature is insufficient to revert the system to its initial hydrophobic state, indicating a high energy barrier for the transition. 

To reduce the influence of moisture, the assembly may be performed in a controlled, anhydrous environment (see, e.g., Refs. \cite{cao2015quality, fan2020transfer, masubuchi2022dry, duleba2023inert, wang2023clean}), though complete water elimination remains challenging. However, a structure taken out of the assembly chamber would inevitably absorb water. Ultrathin metallic contacts\cite{xie2017graphene} reduce strain and thus may suppress strain-induced metastability. Alternative approach involves intentional preconditioning through thermal cycling to lock the system into a stable delaminated state. These insights provide guidelines for device engineering.

Our observation discovers a useful experimental handle: the degree of disorder perceived by the electronic system: transport and optical phenomena can be studied before and after the thermal cycling and different results are expected on the same sample.
These findings not only present new challenges for 2D electronics but also demand a critical re-examination of the prior studies. It appears that the breakdown of the Van der Waals bond is an overlooked yet important parameter.

\section{Conclusions}

In summary, we performed repeated cryogenic transport measurements on graphene/hBN heterostructures transferred onto arrays of metallic islands and contact electrodes. Initial cooldown experiments revealed strong van der Waals adhesion between the graphene and the islands, as evidenced by the gate voltage and magnetic field dependencies of the resistance. However, subsequent thermal cycling between room and cryogenic temperatures disrupts this interfacial bonding, leading to markedly different transport properties characterized by enhanced uniformity and disorder. Optical microscopy confirmed structural changes after the cooldown. Atomic force microscopy provided evidences for thermal-cycling-driven delamination of the hBN/graphene heterostructure.
Raman scattering confirms the existence of local strain in the structure. Hot pressing at elevated temperatures restores van der Waals adhesion. The presented observation allow to attribute the metastability to the breakdown of van-der Waals bonds due to thermal expansion. Thermally activated redistribution of interfacial water layers is evidenced from transport experiments and likely promotes the delamination. Beyond its fundamental implications, this controllable metastability opens new routes for engineering 2D systems on structured surfaces, with potential applications ranging from reconfigurable electronics to strain-sensitive quantum devices.

\acknowledgments{This work was supported by the Russian Science Foundation under Grant No. 23-12-00340. Sample fabrication was conducted at the Shared Facility Center of the P.N. Lebedev Physical Institute.}

\bibliography{ref}

\end{document}